# Silicon photonics LMA amplifiers: High power, high gain, low noise and tunable polarization sensitivity

JAN LORENZEN[1,#], NEETESH SINGH[1,*], KAI WANG[2], SONIA M. GARCIA-BLANCO[2] AND FRANZ X. KÄRTNER[1,3]

[1]*Center for Free-Electron Laser Science CFEL, Deutsches Elektronen-Synchrotron DESY, Germany*
[2]*Integrated Optical Systems, MESA+ Institute for Nanotechnology, University of Twente, 7500AE, Enschede, The Netherlands*
[3]*Department of Physics, Universität Hamburg, Jungiusstr. 9, 20355 Hamburg, Germany*
[#] *jan.lorenzen@desy.de*, [*] *neetesh.singh@desy.de*

**Abstract:** High-power amplifiers are of great importance in many optical systems deployed in optical sensing, ranging, medical surgery, material processing and more. Likewise, high-gain, low-noise amplifiers with low polarization dependence are critical components of long-range optical communication systems. Integrated photonic solutions show great potential in challenging application fields thanks to their drastic reduction in size, weight and cost, but this comes at the expense of low optical power due to reduced energy storage capacity in small devices. Recently, the large mode area (LMA) technology, which is well known for dramatically increasing the output power of fiber amplifiers by orders of magnitude, has been brought to the chip-level. With a large optical mode in the gain medium, the energy storage capacity and saturation power are increased significantly, allowing for high-power amplification with watt-level output power directly from the chip. In this work we demonstrate that a single integrated LMA amplifier is capable of both high-power amplification up to 800 mW with output saturation powers > 115 mW as well as high small-signal net gain up to 30 dB and low-noise amplification with noise figures < 4 dB. The LMA design further allows for a tuning of the polarization dependent gain (PDG) by adjusting the pump power and pump polarization, making it possible to completely nullify the polarization-sensitivity for any given signal power from the µW to mW-level. The power and noise performance achieved surpasses the performance level of many integrated amplifiers and fiber-based amplifiers. We believe that the tunability of the PDG combined with high gain and low noise figure can play a disruptive role for next-generation integrated amplifiers in telecommunication networks.

## 1. Introduction

There are important attributes that one desires of an optical amplifier such as high net gain, high saturation power, low noise amplification and polarization insensitivity. High power is demanded in many applications ranging from optical sensing, detection and ranging, medical surgery, material processing, amplification of optical frequency combs, mode-locked lasers and more [1–10]. Low noise and low polarization sensitivity are crucial for long-range optical telecommunication to ensure high average channel capacity [11]. Fiber-based amplifier systems possess these qualities, but they tend to be large in size and therefore are not compliant with the growing interest in miniaturized and mass-producible systems [12–14], and are not suitable for applications in challenging environments such as deep space [15–18], which require compact form factors. Advancements in chip-scale semiconductor optical amplifiers (SOA), that are hybridly or heterogeneously integrated to a silicon photonics platform, show great potential due to the ease of electrical pumping, high optical gain, and advancements in fabrication yield and integration techniques [19–23]. However, the gain saturation power is usually limited to a few tens of mW [20,21,24–27], especially in the 2 µm wavelength region [26], where the output power is typically much lower than in the C-band. Furthermore, low-noise amplification with integrated SOAs is challenging [25], and the gain can be quite sensitive to the state of polarization of the signal (which is also known as polarization dependent gain (PDG)) [24], even in non-integrated SOAs [28]. Low-PDG amplifiers are especially relevant in long-haul communication systems as the effect accumulates over multiple amplifiers and decreases the average channel capacity [11].

Another class of integrated amplifiers are based on a rare earth-doped (RE) gain media [29–45]. The interest in integrated RE-based amplifiers has risen, even if they are optically pumped, due to the superior performance they offer over their semiconductor counterpart with regards to, for example, noise, saturation power, optical nonlinearity and thermal instability. There are various demonstrations of RE-based integrated amplifiers in the last decades; however, due the tight mode confinement, the saturation power remained below 30 mW and the on-chip output power has been limited to less than 150 mW [44]. Only recently high-power amplification has been shown in silicon photonics with the help of the large mode area (LMA) technology [46–48], in which the optical mode area on-chip is enhanced by a factor of 30 or more compared to standard silicon photonics, while maintaining a high overlap with the gain medium > 90 %. Therefore, the LMA waveguide design allows the light to interact with a large number of gain ions within a short length, which significantly increases the saturation power and energy storage capacity, while supporting only fundamental mode propagation. In recent works, the devices were designed for the power amplification regime to amplify seed signals of several milliwatts of power, which is the power level usually associated with integrated lasers, to well over a watt output power on-chip. In these cases, the maximum net gain was limited to 16 dB before parasitic lasing from the facet reflections kicked in.

In this work we demonstrate LMA-based amplifiers with high small-signal net gain reaching up to 30 dB and a high output saturation power of more than 100 mW, thus being capable of both high-gain operation and high-power amplification within a single device. Additionally, the lowest noise figure was measured to be around 3.6 dB at 30 dB gain. Furthermore, we demonstrate a tunable PDG performance with the capability to minimize it to 0 dB. We show 0 dB PDG over a large range of signal power levels (from a few microwatts to several milliwatts) by tuning either the power or the polarization state of the pump launched into the amplifier. In the following, section 2 describes the device design and section 3 contains the results on high net gain, noise figure and the concept and results on PDG, which is followed by discussion and conclusion.

## 2. Integrated Amplifier Design

A cross-section of the 10.7 cm-long amplifier waveguide layer stack is shown in Fig. 1a) and a top view of the waveguide layout is shown in Fig. 1b). The waveguide consists of a silicon nitride (SiN) layer buried in silica ($SiO_2$) and a RF-sputtered thulium-doped aluminum oxide top layer ($Al_2O_3$:$Tm^{3+}$), which provides the gain. Exact dimensions of the layer stack are listed in the methods section. Although mainly erbium-doping is of interest for telecom applications, we have chosen thulium-doping due to its various medical, defense and space applications as well as possible applications in telecommunication at 2 µm exploiting its large gain bandwidth [49–51]. The estimated thulium concentration and passive film loss are $4.0 \times 10^{20}$ $cm^{-3}$ and ≤ 0.10 dB/cm at 1.61 µm, respectively. The serpentine amplifier waveguide combines high-confinement sections at the input, the output and the bends with long and straight LMA sections highlighted by the blue box in Fig. 1b). The transition to high-confinement elements is realized with adiabatic tapers, allowing for tight bends and a small device footprint < 11 $mm^2$ as well as the potential for seamless integration with other silicon photonic components. In the LMA sections, the TE-polarized pump (1.61 µm) and signal (~ 1.85 µm) light propagate mostly in the gain layer with effective mode areas of 23 and 21 $µm^2$, respectively. The mode profiles in both sections were measured with an IR camera and are shown in Fig. 1c) (see methods for details). This device was designed for operation with the fundamental TE mode, because the TM mode is significantly smaller (~ 9 µm²) and has much more overlap with the rough sidewalls of the SiN waveguide below the gain film leading to higher propagation loss (0.2 % overlap with the SiN for TE and 2.2 % for TM), which means that the overall propagation loss in the LMA section is slightly higher than the pure $Al_2O_3$ film loss. An intraband pumping scheme is used with the pump wavelength at 1.61 µm and the broad emission of the $Tm^{3+}$-doped $Al_2O_3$ provides gain from 1.7 to 2.1 µm. A simplified energy level diagram of the $Tm^{3+}$ ions is shown in Fig. 1d), highlighting also the energy-transfer upconversion process (ETU) as a parasitic ion-ion-interaction process [52,53], which can significantly decrease the excited state lifetime and degrade the amplifier performance [54–56,36,57–60]. In our measurements we have observed the presence of ETU and very small amounts of concentration quenching, although the high-gain benefits of a higher doping concentration far outweighed the negative side effects of ETU and quenching.

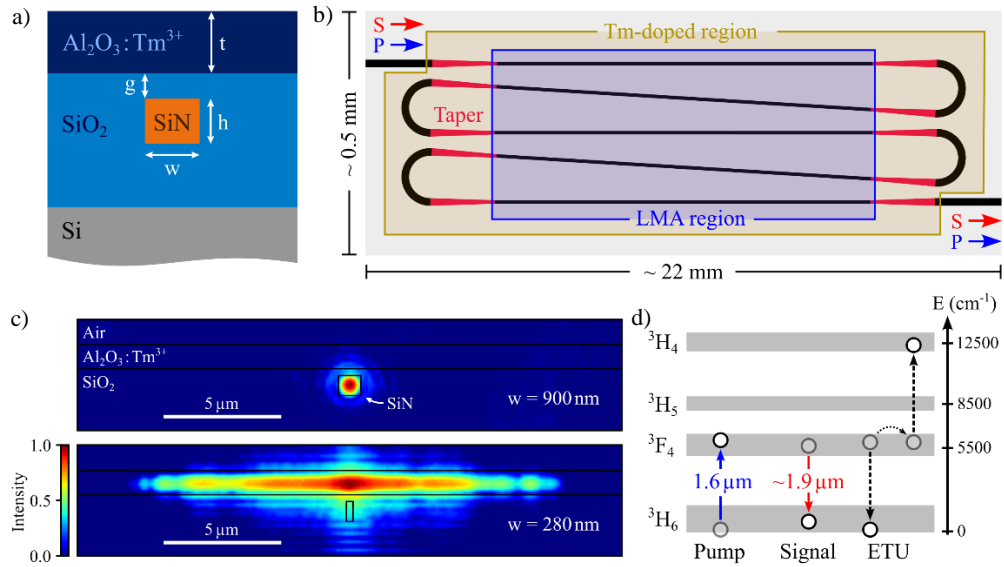

Fig. 1. a) Cross-section of the chip. b) Schematic of the on-chip amplifier. c) Images of the mode profiles in the confined and LMA region at 1850 nm wavelength measured with an IR camera. d) Simplified energy level diagram of $Tm^{3+}$ (ETU: energy-transfer upconversion).

## 3. Results

### 3.1 Gain Measurements

Two sets of gain measurements were performed and the measurement setup is shown in Fig. 2a). A counterpropagating pump setup was chosen to decouple pump and signal input. First, high-power amplification tests were carried out with high-power signals at various wavelengths to investigate the power-handling capability and the gain bandwidth of the amplifier. The second set of measurements were performed with low-power signals at a fixed wavelength of 1818 nm. To avoid facet reflection-based parasitic lasing in the amplifier at high gain (> 16 dB), a high-power index-matching fluid (Norland NOA148, n ~ 1.46 at 1818 nm) was applied to the waveguide facets to reduce facet reflections to roughly -35 dB. It should be noted that such reflections can also be avoided by using angled taper couplers instead of straight couplers. To accurately determine the on-chip net gain, the signal output power of the 10.7 cm long amplifier waveguide was compared to the output power of a 2.2 cm-long low-loss passive straight reference waveguide on the same chip (propagation loss in the high-confinement SiN waveguide < 0.15 dB/cm). As the taper couplers in both waveguides are identical and the rest of the setup remains unchanged, this method directly provides an accurate value of the on-chip net gain, which is independent of coupling and fiber component losses down the line (i.e. WDM and splitter losses). The on-chip signal power was subsequently calculated using the measured coupling loss (see methods). In this report, the depicted and discussed power levels are always the on-chip power levels, unless stated otherwise.

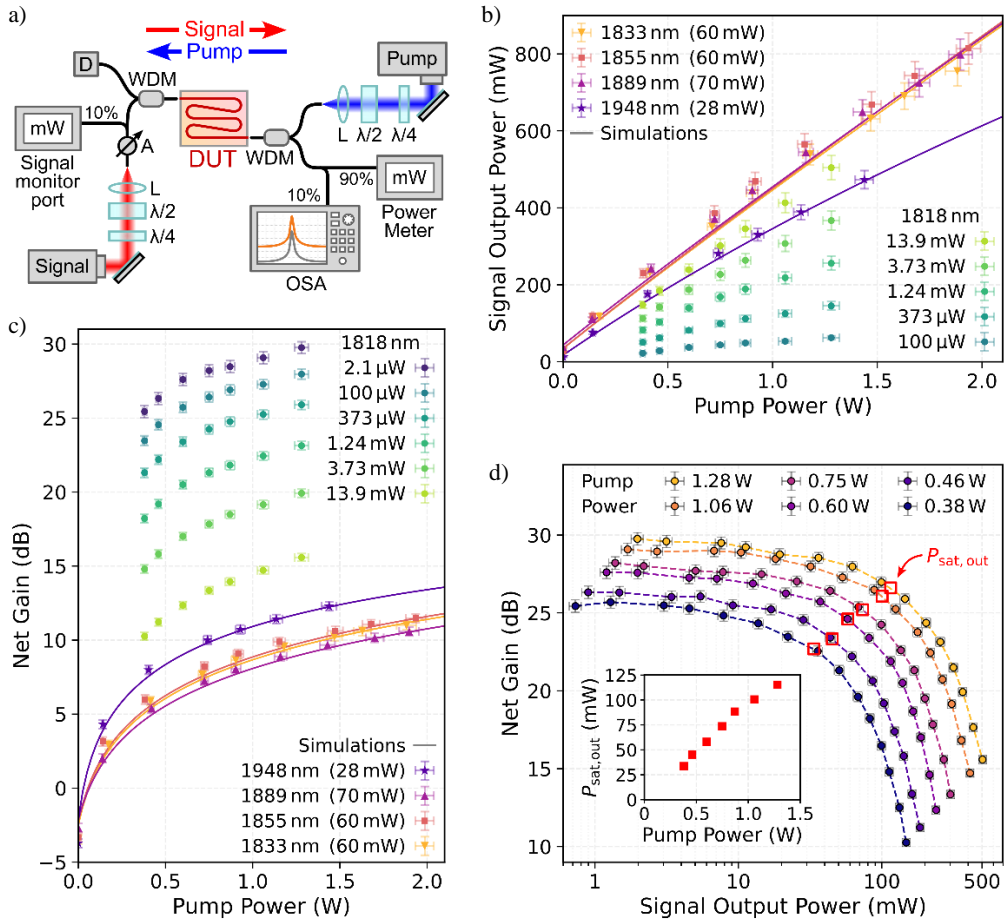

Fig. 2. a) Gain measurement setup with a counter-propagating pump-scheme. Amplified on-chip output power (b) and on-chip net gain (c) as a function of pump power for various signal input power levels (symbols). Solid lines show simulated results for the high-power signals. d) On-chip net gain versus signal output power for different pump power levels. Dashed lines are a guide for the eye. Stars indicate the output saturation power $P_{sat,out}$ and the inset shows $P_{sat,out}$ as a function of pump power.

The measured amplified signal power at the output and the on-chip net gain are shown in Fig. 2b) and c), respectively, as functions of pump power for various signal input power levels and wavelengths. Output powers reached up to 815 mW with high-power input signals from 60 to 70 mW at 1.9 W pump power. The power conversion efficiency (signal output power divided by launched pump power) at the maximum signal output is around 43 %. High-power amplification to > 500 mW was achieved over a broad bandwidth from 1818 to 1948 nm, owing to the broad and nearly flat emission spectrum of Thulium. Signs of gain saturation with increasing pump were only noticeable in the 1948 nm signal, because this wavelength is further away from the gain maximum of thulium (1830 - 1880 nm). Still, a net gain of 12.4 dB was achieved for a 28-mW input signal at this wavelength, before parasitic lasing from the waveguide facets set in and suppressed any further gain at higher pump power levels. Measurements of the thulium emission spectrum further indicate that amplification at even longer wavelengths around 2 µm is possible, but may require small adjustments to the amplifier designs, for example longer gain waveguides to compensate for the smaller emission cross-section at long wavelengths. Simulated amplifier results are shown as solid lines in Fig. 2b) and c) and are based on waveguide and mode parameters as well as measured spectroscopic properties, such as the excited state lifetime, emission and absorption cross-sections and upconversion parameters.

For the small-signal gain measurements at 1818 nm wavelength, index-matching glue was applied to the waveguide facets to prevent parasitic lasing, and the results are also shown in Fig. 2b) and c). Signals with input powers ranging from 2.1 µW to 13.9 mW were tested and a maximum on-chip net gain of 29.8 dB was measured for the lowest input power at 1.28 W pump power before parasitic lasing in the amplifier set in. As the signal input power is increased, the achievable net gain begins to saturate (also known as gain compression), because the strong signals significantly deplete the population inversion. This is highlighted in Fig. 2d), where we show the net gain as a function of signal output power. Initially, the small-signal gain is nearly constant for signal output powers up to ~ 10 mW, after which the gain begins to drop noticeably. From this data we extracted the output saturation power $P_{\text{sat,out}}$, which is taken as the signal output power at which the net gain has dropped by 3 dB from its small-signal gain value. It increases roughly linearly with pump power and reaches slightly more than 115 mW at the highest tested pump power. The high output saturation power can be attributed to the LMA waveguide design, which significantly increases the intrinsic saturation power ($P_{\text{sat}} \sim A_{\text{eff}}$) compared to high-confinement silicon photonics and allows for the amplification of very high-power signals.

## 3.2 Noise Figure

The amplified output signal is accompanied by a broad background of amplified spontaneous emission (ASE) as shown in the spectra in Fig. 3a). ASE noise is the main contributor to the noise of optical amplifiers. To quantify the noise properties, we determined the noise figure (NF) from the measured output spectra using the relation $\text{NF} = P_{\text{ASE}} / (G_{\text{lin}} h\nu B_0) + 1/G_{\text{lin}}$ [61], in which $P_{\text{ASE}}$ is the ASE power within the calibrated OSA equivalent noise bandwidth $B_0 = 0.037$ nm, $h\nu$ is the signal photon energy and $G_{\text{lin}}$ is the gain factor in linear scale (output power divided by input power). The ASE power at the signal wavelength was estimated by measuring it 4 nm away from the signal, which is a good approximation as the ASE spectrum is very flat around the signal.

The measured noise figure values are shown in Fig. 4b) and c) highlighting the dependency on signal input power and on-chip net gain. Values as low as 3.7 dB were measured for small signals < 10 µW. This is the result of the high gain (> 25 dB) and a low spontaneous emission factor $n_{\text{sp}} \propto \text{NF}$, which is given by $n_{\text{sp}} = \sigma_{\text{em}} N_1 / (\sigma_{\text{em}} N_1 - \sigma_{\text{abs}} N_0)$, in which $N_0$ and $N_1$ are the ground state and excited state populations, and $\sigma_{\text{em}}$ and $\sigma_{\text{abs}}$ are the emission and absorption cross-sections. The spontaneous emission factor is low, because the amplifier is pumped very strongly and is thus well inverted ($N_1 \gg N_0$) throughout most of its length, and the weak signals are not strong enough to significantly reduce the population inversion. With higher input power > 100 µW, the amplified signal gets strong enough to deplete the population inversion towards the output section of the amplifier, which reduces the extractable gain and increases the reabsorption loss leading to a higher noise figure [61]. The optimal noise figure in the high-gain regime may be estimated via $\text{NF} \approx 2 n_{\text{sp}}$. Using measured cross-section values we calculate a lower limit of $\text{NF} \approx 2.083 = 3.18$ dB. The reason for the low noise figure of inband-pumped thulium-based amplifiers lies in the low spectral overlap of the pump and emission bands. This results in large differences between $\sigma_{\text{em}}$ and $\sigma_{\text{abs}}$ at the pump wavelength, which enables a high population inversion as $N_1 / (N_0 + N_1) \approx \sigma_{abs} / (\sigma_{em} + \sigma_{abs})$, and at the signal wavelengths, leading to the low $n_{sp}$ and low noise figures very close to the high-gain signal-spontaneous noise quantum limit of 3 dB have been demonstrated with TDFA [62]. This is for example very challenging with inband-pumped erbium-based amplifiers, as the emission and absorption bands have a much stronger overlap. Therefore, $\sigma_{em}$ and $\sigma_{abs}$ are not very different at the usual pump (1460 – 1490 nm) and signal wavelengths (~ 1550 nm) and noise figures are typically noticeably higher than the 3 dB limit [63]. Improved noise figures can be achieved by pumping the erbium ions to a higher excited state using for example a 980 nm pump laser (or a 790 nm pump laser in the thulium case), which allows for an inversion approaching unity and noise figures very close to 3 dB [63,64].

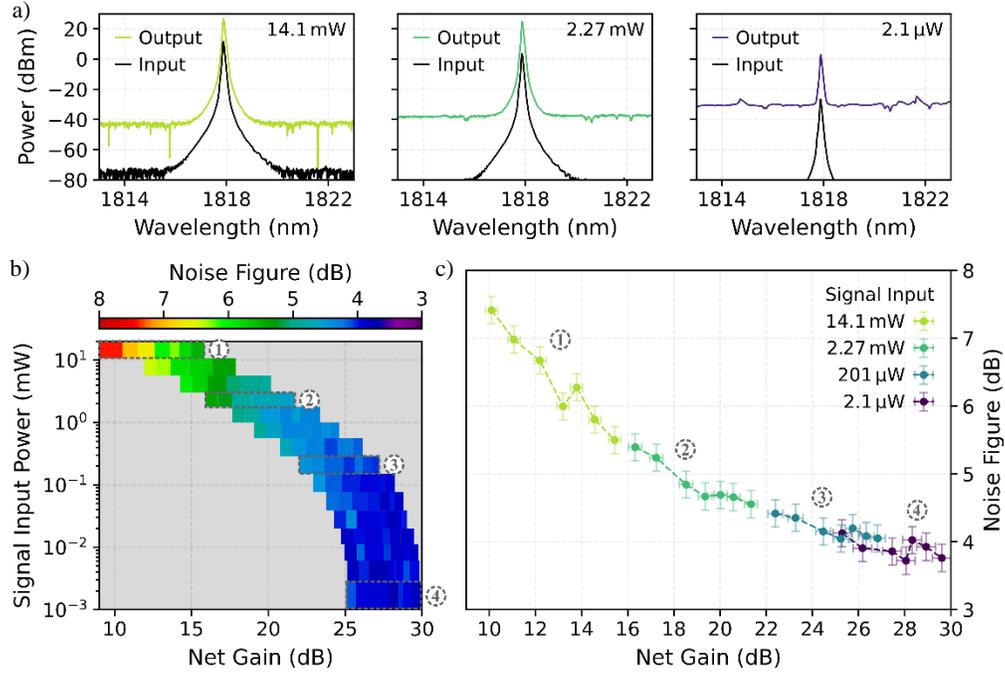

Fig.3. a) Input and output signal spectra for various signal input power levels. b) Colormap of measured noise figure values as functions of net gain and signal input power. c) Noise figure values as a function of net gain for selected signal input power levels.

### 3.3 Tunable Polarization Sensitivity

A second amplifier device was fabricated to demonstrate tunable polarization dependence with the integrated LMA technology. Here we define the polarization dependent gain (PDG) as the gain difference between a TE signal and a TM signal, PDG = $G_{TE} - G_{TM}$, specifying that positive PDG values indicate stronger TE gain, while negative values indicate stronger TM gain. The device was fabricated with a slightly thicker gain layer and a protective $SiO_2$ top cladding leading to large mode areas for both the TE and TM polarized fundamental modes (TE: 57 µm$^2$, TM: 26 µm$^2$) and thus high gain for both polarizations. Simulated mode profiles are shown in Fig. 4a). The TE mode experiences slightly more propagation loss than the TM mode mostly due to a bigger overlap with the lossy $SiO_2$ top cladding and to a lesser extent due to a higher probability of interacting with any defects in the gain layer because of the larger mode, which can be improved with film deposition by reducing film stress and micro cracks. Due to the different mode sizes, the TE and TM signal gain are not generally identical but can be equalized to achieve polarization-independent gain (PDG of exactly 0 dB) by tuning the pump parameters, specifically the pump power and pump polarization. Tuning of the pump conditions changes the spatial overlap of the pump mode with the TE and TM signal modes, since the two are of very different mode size, leading to a differential change in gain for the two polarizations.

To investigate this polarization-dependent behavior, we measured the net gain of TE and TM signals at three signal power levels ranging from 10 µW to 15 mW with TE polarized pump light, and the data is shown in Fig. 4b). The same measurements were also repeated with TM polarized pump light but are not shown here for clarity in the graphs. Initially at low pump power, the TM signal gain is stronger than the TE gain, mainly due to the slightly higher loss with the larger TE mode. Additionally, the smaller TM mode overlaps mostly with the high-intensity parts of the TE pump mode where the population inversion is highest, while the TE signal mode also overlaps with the low-intensity wings of the pump mode, which leads to more re-absorption loss for the TE signal. However, as the pump power is increased the gain medium is well inverted even in the wings of the pump mode and the large TE mode effectively interacts with more excited ions to extract more gain than the TM mode. This behavior has a noticeable dependency on the signal power, because the point where the TE gain overtakes the TM gain shifts to higher pump power when the signal power is lowered, as highlighted in Fig. 4b). This is the result of the higher saturation power of the TE signal due to the larger mode area ($P_{sat} \sim A_{eff}$), meaning that a high-power TE signal can still extract more gain while the TM signal is already saturated (gain compression). This is not the case with low signal power where both the TE and TM signal are far from compression and therefore the TM gain remains high due to the advantages mentioned above. As a result of the crossing of the TE and TM gain curves, one can find a specific pump power for any given signal power at which the PDG is exactly 0 dB, and the determination of this point is only limited by the accuracy of the gain measurement. For example, with this device the 0 dB-point is around 0.66 W pump power for a high-power 15 mW signal, which can be tuned to 0.86 W pump power for an intermediate signal power of 0.2 mW and may be further tuned to approximately ~ 1.03 W pump power for a low-power 10 µW signal. We could not confirm the small-signal 0 dB-point directly by measurement, because we avoided pump powers > 0.9 W since the index-matching fluid on the waveguide facets began to degrade from the prolonged high-power pumping, which can be circumvented with optimized high-power index-matching glue, angled inverse taper couplers or antireflective coating. Nonetheless, from the measured data up to this pump level the 0 dB-point can be extrapolated.

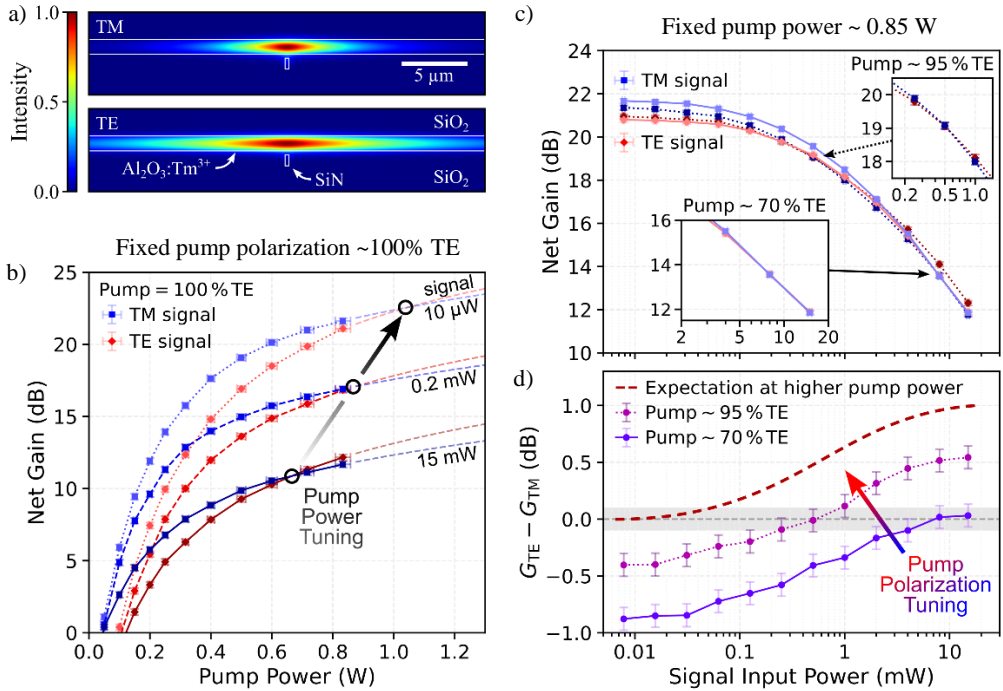

Fig.4. a) Mode profiles of the TM and TE mode in the LMA gain section of the tunable-PDG-amplifier. b) Measured on-chip net gain of TE (red, diamonds) and TM (blue, squares) signals as a function of coupled pump power for three signal input powers. In this measurement the pump light was fully TE polarized. Faint dashed lines are extrapolations of the gain data at higher pump power. 0 dB-PDG points are marked with circles. c) Measured on-chip net gain of TE (red, diamonds) and TM (blue, squares) signals as a function of signal input power with two different mixed pump polarizations (solid lines: 95% TE / 5% TM; dotted lines: 70% TE / 30 % TM). Here the pump power was fixed at 0.85 W on-chip. d) Difference between TE and TM signal gain (PDG) corresponding to the data with mixed polarizations from c). The dashed line is the expected PDG curve with 100% TE pump polarization and approximately 20% more pump power, to achieve 0 dB gain difference at low signal power, which was extrapolated from the data in b).

Alternatively, with the same device, the PDG can also be tuned by changing the polarization state of the pump light while keeping the pump power fixed. We demonstrated this by measuring the TE and TM signal gain with a fixed pump power (0.85 W on-chip) and two mixed polarization states of the pump. In these measurements we explicitly tested the dependence on the signal power ranging from 8 µW to 15 mW and the results are shown in Fig. 4c) and d). The first pump polarization state was mostly TE-polarized (~ 95%), which was optimized for 0 dB PDG around 0.5 mW signal power. For the second pump state we tuned the polarization more towards TM (~ 70% TE, 30% TM) for an optimized PDG at 10 mW signal power. The shift of the PDG with varying pump polarization is the result of the spatial overlap of pump and signal modes. The best gain for each signal polarization is achieved when pump and signal are in the same polarization as then both modes are of similar size and overlap well, while the gain is lowest when pump and signal are in orthogonal polarizations. This effect can be used to lower the gain of one signal polarization while simultaneously increasing the gain for the other. For example, with this device at 0.85 W pump power and TE pump polarization, the high-power signal gain is slightly stronger in TE than TM signal polarization. To equalize the gain, the pump can be tuned to be partially TM polarized, which effectively reduces the pump mode overlap with the TE signal and improves the overlap with the TM signal, thus bringing the gain for both polarizations closer together. This is highlighted in Fig. 4d), demonstrating how the PDG can be optimized for various signal powers by tuning the pump polarization. Achieving a PDG of 0 dB at low signal powers ~ 10 µW with this device would require approximately 20% more pump power, as extrapolated from the data in Fig. 4b).

When operating in the high-gain regime (> 0.8 W pump power in this case), which is typically desired for the benefits of low noise, the PDG stays below 1 dB over a large range of signal powers from 8 µW to 15 mW (> 30 dB range). After optimization via pump power or polarization tuning, the PDG is robust against small signal power fluctuations with PDG variations of less than ± 0.1 dB within a ~ 6 dB range of signal powers. If required, the PDG may also be increased significantly by lowering the pump power and tuning the polarization to mostly TM. In this case, depending on the signal power, the TM signal gain can be up to 8 dB stronger than the TE gain. While tuning of the pump power can be realized rather effortlessly in a fully integrated amplifier device, e.g. with hybridly integrated pump laser diodes, the polarization tuning on-chip remains challenging. An approach may be to integrate two pump diodes with orthogonal polarization in a bi-directional pump scheme, such that the overall pump polarization state can be tuned through the power of each diode.

## 4. Discussion

The results show that a single integrated LMA amplifier is suitable for both small-signal amplification with high net gain and low noise figures as well as high-power amplification with signal powers approaching the watt-level, which was previously not achievable with high-confinement silicon photonics amplifiers. The broad emission spectrum of the thulium gain ions enables very broadband amplification up to 2 µm in wavelength, making the amplifier suitable for multiple applications in gas sensing and medical fields due to absorption lines of several molecules in this spectral range, such as $H_2O$, $CO_2$ or $NH_3$. Output saturation powers > 115 mW have been demonstrated with this device, which is almost one order of magnitude higher than any other integrated amplifier to this date. Higher saturation powers can be achieved by increasing the mode area even further with a thicker gain layer and thicker interlayer oxide (dimensions t and g in Fig. 1a), and mode areas > 100 µm$^2$ are achievable. Such large mode areas are beneficial for high-power amplification within a short length as in power amplifier operation, but are less desirable for small-signal amplification, where longer devices are needed for the gain to build up. The power conversion efficiency with the 10.7 cm-long amplifier tested here was limited to 43 %, which can be improved to > 50% and even ~ 60% by reducing the total propagation losses and optimizing the length and doping concentration, as we have recently demonstrated with a short, high-concentration device [46,47]. Despite the high thulium concentration, the effects of energy-transfer upconversion and concentration quenching had only a negligible impact on the amplifier performance, implying that the concentration may be further increased to improve the gain even further. This also suggests that a high doping concentration may be employed in future erbium-doped devices without sacrificing amplifier performance, despite the higher tendency of erbium to form clusters leading to quenching and amplifier degradation [36,56].The low noise figure and tunable PDG are interesting for applications in telecommunication where currently erbium-doped fiber amplifiers (EDFA) deliver high-quality signal amplification but are envisioned to be replaced by more power and space efficient chip-scale devices. The current amplifier is fabricated with thulium doping, but future devices will also employ erbium-doping, which has already been demonstrated to great success in the aluminum oxide host material [45]. Combined with the LMA waveguide architecture and further design optimization, we expect similar if not better high-power and low-noise amplification in the C-band that is on par with EDFA in performance, while reducing the device footprint by several orders of magnitude. Our proof-of-concept measurements demonstrate the ability of the LMA amplifier to tune the PDG to exactly 0 dB for a wide range of signal power levels by modifying the pump power or pump polarization even when the modes are of different sizes (unlike in fiber amplifiers and SOA). This is a big challenge in integrated SOAs and high-confinement RE-based amplifiers, where the PDG is fixed for a given device geometry and pump conditions, and the PDG cannot be easily adjusted once fabricated. Further tests are planned to explore the effects of polarization hole burning (PHB) in the LMA amplifier, which is a main driving mechanism behind non-zero PDG in EDFA chains [65,66]. With an adjusted design the PDG can also be increased significantly to suppress one signal polarization compared to the other. For example, with a thinner gain film the TM mode has a stronger overlap with the rough SiN waveguides, which leads to much higher propagation loss compared to the TE mode. This can lead to a PDG > 15 dB, which effectively turns the amplifier into a polarizing element heavily favoring TE mode propagation and gain.

## 5. Conclusion

In conclusion, we demonstrate high net gain up to 30 dB and high output saturation power > 115 mW in an integrated thulium-doped amplifier based on large mode area technology. The amplifier exhibits a low noise figure down to 3.7 dB due to the high gain and high level of inversion achievable with intraband-pumped thulium. Furthermore, we demonstrate low polarization dependent gain, which is tunable over a signal power range of more than 30 dB by tuning the pump power or pump polarization. This enables the tuning of the PDG to exactly 0 dB and even with fixed pump parameters the PDG stays below 0.1 dB over a 6 dB range of signal powers, which opens up possibilities for the implementation of low-noise integrated amplifiers in long-range telecommunication networks.

## 6. Methods

The dimensions of the chip layer stack shown in Fig. 1a) are h = 800 nm, t = 1000 nm, and g = 300 nm. The waveguide width in the high-confinement sections is w = 900 nm and in the LMA section it is w = 280 nm. In the device used for the PDG characterization the gain layer was slightly thicker at t = 1100 µm and a 1 µm-thick SiO2 top cladding layer was added. The images of the modes shown in Fig. 1c) in the LMA and confined sections were captured with two aspheric lenses, one with 8 mm focal length to collimate the waveguide output and the other with 300 mm focal length to create the image on an infrared camera (ICI SWIR 320).

The gain measurement setup (see Fig. 1a) consists of a high-power CW pump laser at 1609 nm wavelength and a homebuilt high-power tunable CW seed laser based on a Tm-doped YLF crystal, tunable from 1830 to 1950 nm via a birefringent filter (reference [67] for details). Alternatively, an interference-based filter was used to reach shorter wavelengths (1818 nm). Signal and pump light polarizations were controlled via half- and quarter-wave plates before coupling into fiber-based wavelength division multiplexers (WDM) to obtain the desired polarization on-chip. The signal output power was monitored on a power meter through the 90% port of a fiber-based power splitter and the amplified spectra were recorded on a calibrated optical spectrum analyzer (OSA) through the 10% port. The fiber-to-chip coupling loss was determined from the total insertion loss of a 2.2 cm long passive waveguide next to the amplifier waveguide on the same chip. For the high-gain device the coupling losses were 2.6 dB per facet for the pump and 3.9 – 4.2 dB for the signal light (1830 – 1950 nm) in TE polarization. With the index-matching glue (Norland NOA 148) applied, the coupling losses were measured to be 2.0 dB for the pump and 2.2 dB for the signal light. For the tunable PDG device a different index-matching glue was applied (Luvantix SH-548HT) and the coupling losses were 4.7 dB and 7.1 dB for the TE and TM polarized pump light, and 3.9 and 4.9 dB for the TE and TM polarized signal light, respectively.

## Funding

This work is supported by EU Horizon 2020 Framework Programme—Grant Agreement No.: 965124 (FEMTOCHIP), and Deutsche Forschungsgemeinschaft (SP2111) contract number 403188360.

## Acknowledgements

We thank Muharrem Kilinc, Mikhail Pergament and Ümit Demirbas for the development of the tunable high-power seed laser, which was used for the high-power gain measurements. We would furthermore like to thank Milan Sinobad for the fruitful discussions.

## Author contributions

J.L. performed the experiment, analyzed the data and wrote the manuscript. NS conceived the work presented, the idea of PDG with LMA waveguide, co-supervised and helped in the experiment and analyzing the data. K.W. and S.M.G.B. deposited the gain film. F.X.K. supervised the project. All authors helped in writing the manuscript.